\documentclass[aps,prd,a4paper,preprint,preprintnumbers,
nofootinbib,superscriptaddress]{revtex4-2}

\usepackage[a4paper, hdivide={1.91cm,,1.165cm},
vdivide={1.83cm,,3.0cm}]{geometry}

\usepackage{xcolor}

\usepackage[hyperfootnotes=false]{hyperref}
\hypersetup{linkcolor=red,
citecolor=green,
filecolor=cyan,
urlcolor=magenta}

\usepackage{amsfonts}

\usepackage{amsmath}
\usepackage{amssymb}

\usepackage{bbm}

\usepackage{bbold}

\usepackage{adjustbox}
\usepackage{booktabs}

\usepackage{colortbl}

\usepackage{cancel}

\usepackage{color}
\usepackage{xcolor}

\usepackage{graphicx}
\usepackage{xspace}

\usepackage{units}

\usepackage{slashed}

\usepackage{tensor}

\usepackage{cleveref}

\usepackage{tabularx}
\usepackage{multirow}

\usepackage{setspace}
\usepackage{orcidlink}

\usepackage{caption}
\usepackage{subcaption}

\usepackage{enumerate}
\usepackage{enumitem}

\usepackage[version=3]{mhchem}

\usepackage{lipsum}

\newcommand{{\ia} }{{\i}}
\newcommand{{\Ia} }{{\.I}}

\newcommand{\ra}{\rightarrow}
\newcommand{\la}{\leftarrow}

\def\s2tw{{\rm sin ^2 \theta_{W}}}

\def\beq{\begin{equation}}
\def\eeq{\end{equation}}
\def\bea{\begin{eqnarray}}
\def\eea{\end{eqnarray}}
\def\ve{\vert}
\def\vel{\left|}
\def\ver{\right|}
\def\nnb{\nonumber}

\def\nnb{\nonumber}
\def\la{\langle}
\def\ra{\rangle}

\def\es{ &=& }
\def\ar{&+& }
\def\ek{&-& }
\def\cp{&\times&}

\def\bea{\begin{eqnarray}}
\def\eea{\end{eqnarray}}
\def\beeq{\begin{eqnarray}}
\def\eeeq{\end{eqnarray}}
\def\ve{\vert}
\def\vel{\left|}
\def\ver{\right|}
\def\nnb{\nonumber}

\def\rar{\rightarrow}
\def\lrar{\leftrightarrow}  
\def\nnb{\nonumber}
\def\la{\langle}
\def\ra{\rangle}
\def\ba{\begin{array}}
\def\ea{\end{array}}

\def\xis0{{\Xi^{*0}}}

\def\g5{\gamma_5}

\def\es{\!\!\! &=& \!\!\!}

\def\ar{&+& \!\!\!}
\def\ek{&-& \!\!\!}

\def\cp{&\times& \!\!\!}

\def\mcdot{\!\cdot\!}

\allowdisplaybreaks[1]

\begin{document}


\title{QCD sum-rule determination of the axial-vector mixing angle in the \texorpdfstring{\(B_c(1P)\)}{Bc(1P)} sector}

\author{T.~M.~Aliev\,\orcidlink{0000-0001-8400-7370}}
\email{taliev@metu.edu.tr}
\affiliation{Department of Physics, Middle East Technical University,
Ankara, 06800, Turkey}

\author{S.~Bilmis\,\orcidlink{0000-0002-0830-8873}}
\email{sbilmis@metu.edu.tr}
\affiliation{Department of Physics, Middle East Technical University,
Ankara, 06800, Turkey}
\affiliation{TUBITAK ULAKBIM, Ankara, 06510, Turkey}

\author{M.~Savci\,\orcidlink{0000-0002-6221-4595}}
\email{savci@metu.edu.tr}
\affiliation{Department of Physics, Middle East Technical University,
Ankara, 06800, Turkey}

\date{\today}

\makeatletter
\def\frontmatter@abstract@produce{%
  \par
  \addvspace{\frontmatter@preabstractspace}%
  \begingroup
    \dimen@\baselineskip
    \setbox\z@\vtop{\unvcopy\absbox}%
    \advance\dimen@-\ht\z@\advance\dimen@-\prevdepth
    \@ifdim{\dimen@>\z@}{\vskip\dimen@}{}%
  \endgroup
  \begingroup
    \prep@absbox
    \unvbox\absbox
    \post@absbox
  \endgroup
  \@ifx{\@empty\mini@notes}{}{\mini@notes\par}%
  \addvspace\frontmatter@postabstractspace
}%
\makeatother

\def\rar{\rightarrow}
\def\Rar{\Rightarrow}
\def\lrar{\leftrightarrow}
\def\bea{\begin{eqnarray}}
\def\eea{\end{eqnarray}}
\def\ve{\vert}
\def\vel{\left|}
\def\ver{\right|}
\def\nnb{\nonumber}
\def\la{\langle}
\def\ra{\rangle}
\def\ba{\begin{array}}
\def\ea{\end{array}}
\def\es{&=&}
\def\ar{&+&}
\def\ek{&-&}
\def\cp{&\times&}
\def\mcdot{\!\cdot\!}

\begin{abstract}
We determine the mixing angle between the \(1^1P_1\) and \(1^3P_1\)
axial-vector states in the \(B_c(1P)\) sector using QCD sum rules. The
operator product expansion includes the gluon condensates and the four-quark
condensate up to dimension six, and the next-to-leading-order perturbative
corrections to all three correlation functions; for the correlator of the
axial-vector and tensor currents these corrections are calculated here for the
first time. The
analysis of the sum rules leads to
\(\theta_{B_c(1P)}=\)\((46.6\pm0.4)^\circ\) for the mixing
angle between these two states. We also compare our result with theoretical studies
available in the literature.
\end{abstract}

\maketitle

\newpage

\section{Introduction}
\label{sec:intro}

The spectroscopy of heavy hadrons provides an important testing ground for
our understanding of the nonperturbative regime of QCD. Among these systems, the \(B_c\) meson family occupies a special
position. In contrast to charmonium and bottomonium, the \(B_c\) mesons are
composed of two heavy quarks with different flavors, \( \bar b c \). This
asymmetric heavy-heavy structure makes the \(B_c\) spectrum richer than
ordinary quarkonium systems and provides a complementary laboratory for
studying the dynamics of heavy quarks inside hadrons. Since both constituent
quark masses are much larger than the QCD scale, the \(B_c\) system is also
well suited for comparisons among lattice QCD, quark models, nonrelativistic
QCD, and QCD sum-rule calculations
\cite{Davies1996,Mathur2018,Godfrey2004,EichtenQuigg2019}.

Experimentally, the ground-state \(B_c\) meson was first observed by the CDF
Collaboration at the Tevatron \cite{CDF1998PRL,CDF1998PRD}. Since then,
significant progress has been achieved in the study of its excited spectrum.
The first evidence for an excited \(B_c\) state was reported by the ATLAS
Collaboration \cite{ATLAS2014}, and the \(2S\) sector was subsequently resolved
by the CMS and LHCb collaborations \cite{CMS2019,LHCb2019}. The ATLAS
Collaboration has also recently reported the observation of a new state in the
\(B_c^+\gamma\) channel, consistent with the lowest vector \(B_c^{*+}\) state
\cite{ATLAS2026BcStar}. More directly relevant for the present work, the LHCb
Collaboration recently reported the observation of orbitally excited \(B_c^+\)
structures in the \(B_c^+\gamma\) mass spectrum \cite{LHCb2025PRL,LHCb2025PRD}.
These structures are compatible with the lowest \(P\)-wave \(B_c^+\) states and
provide new experimental motivation for refined theoretical studies of the
\(B_c(1P)\) sector.

The \(1P\) \(B_c\) multiplet contains four states: the spin-singlet \(1^1P_1\)
configuration and the spin-triplet \(1^3P_0\), \(1^3P_1\), and
\(1^3P_2\) configurations. Since the \(B_c\) meson is composed of two
different heavy flavors, charge conjugation is not a good quantum number.
Consequently, the two axial-vector configurations \(1^1P_1\) and
\(1^3P_1\) can mix and form the physical \(J^P=1^+\) axial-vector
states. In the following
we use the notation \(B_c(1P)\) to specify the theoretical sector under
study; the individual quantum-number assignments of the
experimentally observed structures have not yet been fully resolved. Following
the convention used in Ref.~\cite{LHCb2025PRD},
\[
\begin{pmatrix}
|1P_1'\rangle \\
|1P_1\rangle
\end{pmatrix}
=
\begin{pmatrix}
\cos\theta & \sin\theta \\
-\sin\theta & \cos\theta
\end{pmatrix}
\begin{pmatrix}
|1^1P_1\rangle \\
|1^3P_1\rangle
\end{pmatrix},
\]
the parameter \(\theta\) denotes the \(1^1P_1-1^3P_1\) mixing angle. This
mixing is not merely a formal property of the spectrum. It affects the
identification of the physical \(B_c(1P)\) states, their radiative transition
patterns, the relative contributions of the different \(P\)-wave states to the
observed \(B_c^+\gamma\) spectrum, and the branching ratios used in
theory-constrained experimental analyses \cite{LHCb2025PRD}.

Several theoretical approaches have been used to study the \(B_c\) spectrum,
including lattice QCD, nonrelativistic and relativistic quark models,
Bethe-Salpeter approaches, Dyson-Schwinger methods, coupled-channel models,
and QCD sum rules
\cite{Davies1996,Gershtein1995,GuptaJohnson1996,Fulcher1999,Ebert2003,
Godfrey2004,EichtenQuigg2019,Li2019,Wang2022,Li2023,LiLiu2023,HaoZhu2024,
Wang2013}. These studies generally predict the \(B_c(1P)\) masses in the
region now probed experimentally. In those analyses where the axial-vector
mixing is discussed, the predicted \(1^1P_1-1^3P_1\) mixing angle shows a
sizable model dependence. For example, the theory-constrained analysis of the
LHCb \(B_c^+\gamma\) spectrum compares predictions in which the mixing angle
ranges from values near \(20^\circ\) to values above \(50^\circ\), depending on
the model input \cite{LHCb2025PRD}. This spread indicates that an independent
determination of the \(B_c(1P)\) axial-vector mixing angle is useful for both
spectroscopy and phenomenological applications.

QCD sum rules provide a complementary nonperturbative framework for relating
hadronic observables to QCD degrees of freedom. In our previous work, this
method was applied to the \(1^3P_1-1^1P_1\) mixing angles of heavy axial-vector
mesons containing one heavy and one light quark \cite{AlievBilmisSavci2024}.
The \(B_c(1P)\) system requires a separate analysis because it contains two
heavy quarks with different flavors.

In the present work, we determine the \(1^1P_1-1^3P_1\) mixing angle of the
axial-vector \(B_c(1P)\) states within the QCD sum-rule approach. Our aim is to
provide an independent estimation of this quantity and to assess
its relevance for the interpretation of the recently observed \(B_c(1P)\)
structures. We include the next-to-leading-order (NLO) perturbative corrections
to the three correlation functions entering the sum rule. For the two diagonal
correlators, these corrections have been calculated in Refs.~\cite{DjouadiGambino1994,GeneretGubernariLoisa2026};
for the mixed correlator of the axial-vector and tensor currents we calculate
them in this work.

The paper is organized as follows: in Sec.~\ref{sec:theory}, we introduce the
interpolating currents, define the correlation functions, and derive the QCD
sum rule for the mixing angle. In Sec.~\ref{sec:numerical}, we present the
numerical analysis of the obtained sum rule and extract the value of the
mixing angle. In this section, we also compare our result with
existing theoretical predictions. The final section is devoted to our
conclusions.

\section{Theoretical framework}
\label{sec:theory}

In this section, we derive the QCD sum rule for the mixing angle between the
\(1^1P_1\) and \(1^3P_1\) axial-vector \(B_c(1P)\) states. We first introduce
the interpolating currents and the corresponding two-point correlation
functions. The correlation functions are represented both at the hadronic
level, in terms of physical intermediate states, and at the QCD level, in the
deep Euclidean region \(p^2 \ll (m_b + m_c)^2\), using the operator product
expansion. Equating these two representations after Borel transformation and
continuum subtraction yields the desired sum rule.

We use the quark-content convention \(B_c^+=\bar b c\). The two
interpolating currents employed for the axial-vector \(B_c(1P)\) system are
chosen as
\begin{align}
J_\mu^A(x) &= \bar b(x)\gamma_\mu\gamma_5 c(x),
\label{eq:current_A} \\
J_\mu^B(x) &= i\,\bar b(x)\sigma_{\mu\alpha}
\frac{p^\alpha}{m_b+m_c}\gamma_5 c(x),
\label{eq:current_B}
\end{align}
where
\begin{equation}
\sigma_{\mu\nu}=\frac{i}{2}[\gamma_\mu,\gamma_\nu].
\end{equation}
The factor \(1/(m_b+m_c)\) in \(J_\mu^B\) is introduced so that the two
currents have the same mass dimension. The corresponding two-point
correlation functions are
\begin{equation}
\Pi_{\mu\nu}^{ij}(p)=
i\int d^4x\, e^{ip\cdot x}
\langle 0 | T\{J_\mu^i(x)J_\nu^{j\dagger}(0)\}|0\rangle ,
\qquad i,j=A,B .
\label{eq:corr_def}
\end{equation}
They are decomposed as
\begin{equation}
\Pi_{\mu\nu}^{ij}(p)=
\left(-g_{\mu\nu}+\frac{p_\mu p_\nu}{p^2}\right)\Pi_1^{ij}(p^2)
+\frac{p_\mu p_\nu}{p^2}\Pi_0^{ij}(p^2),
\label{eq:lorentz_decomp}
\end{equation}
where the transverse amplitude \(\Pi_1^{ij}\) contains the spin-one
contribution. We project it using
\begin{equation}
\Pi_1^{ij}(p^2)=
\frac{1}{3}
\left(g^{\mu\nu}-\frac{p^\mu p^\nu}{p^2}\right)
\Pi_{\mu\nu}^{ij}(p).
\label{eq:projected_corr}
\end{equation}

The physical axial-vector currents are obtained by rotating the current basis,
\begin{equation}
\begin{pmatrix}
J_\mu^{(1)} \\
J_\mu^{(2)}
\end{pmatrix}
=
\begin{pmatrix}
\cos\theta & -\sin\theta \\
\sin\theta & \cos\theta
\end{pmatrix}
\begin{pmatrix}
J_\mu^A \\
J_\mu^B
\end{pmatrix}.
\label{eq:current_rotation}
\end{equation}

{
Here \(J_\mu^A\) and \(J_\mu^B\) are the basis currents associated with the
\(1^3P_1\) and \(1^1P_1\) configurations, respectively. With the convention
for the physical states introduced in Sec.~I, \(J_\mu^{(1)}\) interpolates
\(A_1\equiv |1P_1\rangle\), whereas \(J_\mu^{(2)}\) interpolates
\(A_2\equiv |1P_1'\rangle\).
}

{
Within the two-state description, the correlation matrix is diagonal in the
physical basis. Denoting the correlation matrix in the original current basis
by
}
{
\[
\boldsymbol{\Pi}_1(p^2)=
\begin{pmatrix}
\Pi_1^{AA}(p^2) & \Pi_1^{AB}(p^2)\\
\Pi_1^{BA}(p^2) & \Pi_1^{BB}(p^2)
\end{pmatrix},
\]
}
{
Equation~\eqref{eq:current_rotation} implies
}
{
\[
R(\theta)\,
\boldsymbol{\Pi}_1(p^2)\,
R^T(\theta)
=
\begin{pmatrix}
\Pi_1^{(11)}(p^2) & 0\\
0 & \Pi_1^{(22)}(p^2)
\end{pmatrix},
\]
}
{
where \(R(\theta)\) is the rotation matrix in
Eq.~\eqref{eq:current_rotation}. Using
\(\Pi_1^{AB}=\Pi_1^{BA}\), the vanishing of the nondiagonal element gives
}
\begin{equation}
\Pi_1^{AB}(p^2)\cos 2\theta
+\frac{1}{2}\left[\Pi_1^{AA}(p^2)-\Pi_1^{BB}(p^2)\right]
\sin 2\theta=0 .
\label{eq:mixing_condition}
\end{equation}

{
Equivalently, writing
\(\boldsymbol{\Pi}_1
=R^T(\theta)\boldsymbol{\Pi}_{1,\mathrm{phys}}R(\theta)\)
gives explicitly
}
{
\begin{align*}
\Pi_1^{AA}(p^2)-\Pi_1^{BB}(p^2)
&=
\cos 2\theta\,
\left[\Pi_1^{(11)}(p^2)-\Pi_1^{(22)}(p^2)\right],
\\
2\Pi_1^{AB}(p^2)
&=
-\sin 2\theta\,
\left[\Pi_1^{(11)}(p^2)-\Pi_1^{(22)}(p^2)\right].
\end{align*}
}
{
Thus, the difference
\(\Pi_1^{(11)}-\Pi_1^{(22)}\) is not assumed to vanish.
The two physical diagonal correlators may contain different masses and pole
residues; their difference appears as a common factor and cancels when the
ratio of the above two relations is taken.
}

After Borel transformation and continuum subtraction, the sum rule for the
mixing angle becomes
\begin{equation}
\tan 2\theta =
-\frac{2\Pi_1^{AB}(M^2,s_0)}
{\Pi_1^{AA}(M^2,s_0)-\Pi_1^{BB}(M^2,s_0)} ,
\label{eq:final_theta_sumrule}
\end{equation}
where \(M^2\) is the Borel parameter and \(s_0\) is the continuum threshold.

The QCD side is calculated in the deep Euclidean region using the operator
product expansion. After applying Wick's theorem to the corresponding correlation functions, we obtain
\begin{align}
\Pi_{\mu\nu}^{AA}(p) &=
-i \int \frac{d^4 k}{(2\pi)^4}
\mathrm{Tr}\left[
\gamma_\mu\gamma_5 S_c(k)
\gamma_\nu\gamma_5 S_b(k-p)
\right],
\label{eq:wick_AA}
\\
\Pi_{\mu\nu}^{AB}(p) &=
-i \int \frac{d^4 k}{(2\pi)^4}
\mathrm{Tr}\left[
\gamma_\mu\gamma_5 S_c(k)
i\sigma_{\nu\alpha}\frac{p^\alpha}{m_b+m_c}\gamma_5
S_b(k-p)
\right],
\label{eq:wick_AB}
\\
\Pi_{\mu\nu}^{BB}(p) &=
-i \int \frac{d^4 k}{(2\pi)^4}
\mathrm{Tr}\left[
i\sigma_{\mu\alpha}\frac{p^\alpha}{m_b+m_c}\gamma_5
S_c(k)
i\sigma_{\nu\beta}\frac{p^\beta}{m_b+m_c}\gamma_5
S_b(k-p)
\right].
\label{eq:wick_BB}
\end{align}
Here the trace is taken over Dirac and color indices. The algebraic
manipulations of the Dirac matrices, Lorentz structures, and color factors
were performed with \textsc{FeynCalc}
\cite{Mertig1991,Shtabovenko2016,Shtabovenko2020}.
In our calculations, we use the fixed-point gauge,
\begin{equation}
x^\mu A_\mu(x)=0,
\end{equation}
in which the gauge potential can be expressed in terms of the field-strength
tensor as
\begin{align}
A_\mu(x)&=
\int_0^1 dt\, t\, x^\nu G_{\nu\mu}(tx)
=
\frac{1}{2}x^\nu G_{\nu\mu}(0)
+\frac{1}{3}x^\alpha x^\nu D_\alpha G_{\nu\mu}(0)
+\cdots
\nonumber\\
&=\sum_{r=0}^{\infty}\frac{1}{r!\,(r+2)}\,
x^\nu x^{\alpha_1}\cdots x^{\alpha_r}\,
D_{\alpha_1}\cdots D_{\alpha_r}G_{\nu\mu}(0) .
\label{eq:fixed_point_gauge}
\end{align}
The massive-quark propagator, including the one-gluon, two-gluon, and three-gluon background-field contributions, can be written in momentum space as follows:
\begin{equation}
S_Q(k)=S_Q^{(0)}(k)+S_Q^{(G)}(k)+S_Q^{(GG)}(k)
+S_Q^{(GGG)}(k)+\cdots ,
\qquad Q=b,c ,
\label{eq:heavy_prop_compact}
\end{equation}
where the notation follows the standard background-field expansion
\cite{Shifman1979,Shifman:1978by,Reinders1985,Colangelo2000,Chen:2014tensor}.
Explicitly, the free and one-gluon parts of the heavy-quark propagator are
\begin{equation}
S_Q^{(0)}(k)=\frac{\slashed{k}+m_Q}{k^2-m_Q^2},
\label{eq:heavy_prop_free}
\end{equation}
and
\begin{equation}
S_Q^{(G)}(k)=
-\frac{g_s G_{\alpha\beta}^A T^A}{4}
\frac{
\sigma^{\alpha\beta}(\slashed{k}+m_Q)
+(\slashed{k}+m_Q)\sigma^{\alpha\beta}}
{(k^2-m_Q^2)^2}.
\label{eq:heavy_prop_one_gluon}
\end{equation}
The two-gluon term of the propagator can be expressed as follows~\cite{Reinders1985,Chen:2014tensor}:
\begin{equation}
S_Q^{(GG)}(k)
=
-\frac{g_s^2}{4}
G_{\mu\rho}^A G_{\nu\sigma}^B T^A T^B
\frac{\slashed{k}+m_Q}{(k^2-m_Q^2)^5}
\left(
f^{\mu\rho\nu\sigma}
+f^{\mu\nu\rho\sigma}
+f^{\mu\nu\sigma\rho}
\right),
\label{eq:heavy_prop_two_gluon_compact}
\end{equation}
while the three-gluon term is
\begin{align}
S_Q^{(GGG)}(k)
&=
\frac{g_s^3}{8}
G_{\mu\alpha}^A G_{\nu\beta}^B G_{\rho\gamma}^C
T^A T^B T^C
\frac{\slashed{k}+m_Q}{(k^2-m_Q^2)^7}
\nonumber\\
&\quad\times
\Big(
f^{\mu\alpha\nu\beta\rho\gamma}
+f^{\mu\alpha\nu\rho\beta\gamma}
+f^{\mu\alpha\nu\rho\gamma\beta}
+f^{\mu\nu\alpha\beta\rho\gamma}
+f^{\mu\nu\beta\alpha\rho\gamma}
+f^{\mu\nu\beta\rho\alpha\gamma}
\nonumber\\
&\qquad
+f^{\mu\nu\beta\rho\gamma\alpha}
+f^{\mu\nu\alpha\rho\beta\gamma}
+f^{\mu\nu\alpha\rho\gamma\beta}
+f^{\mu\nu\rho\alpha\beta\gamma}
+f^{\mu\nu\rho\beta\alpha\gamma}
+f^{\mu\nu\rho\beta\gamma\alpha}
\nonumber\\
&\qquad
+f^{\mu\nu\rho\alpha\gamma\beta}
+f^{\mu\nu\rho\gamma\alpha\beta}
+f^{\mu\nu\rho\gamma\beta\alpha}
\Big).
\label{eq:heavy_prop_three_gluon_compact}
\end{align}
The shorthand \(f\) denotes the ordered Dirac string
\begin{equation}
f^{\mu\nu\cdots\alpha\beta}
=
\gamma^\mu(\slashed{k}+m_Q)\gamma^\nu(\slashed{k}+m_Q)
\cdots
\gamma^\alpha(\slashed{k}+m_Q)\gamma^\beta(\slashed{k}+m_Q).
\label{eq:f_definition}
\end{equation}
In these expressions, \(T^A=\lambda^A/2\), where
\(\lambda^A\) are the Gell-Mann matrices in color space, and \(A,B,C\) label
color-adjoint indices.

\begin{figure}[tbp]
\centering
\includegraphics[width=0.67\textwidth]{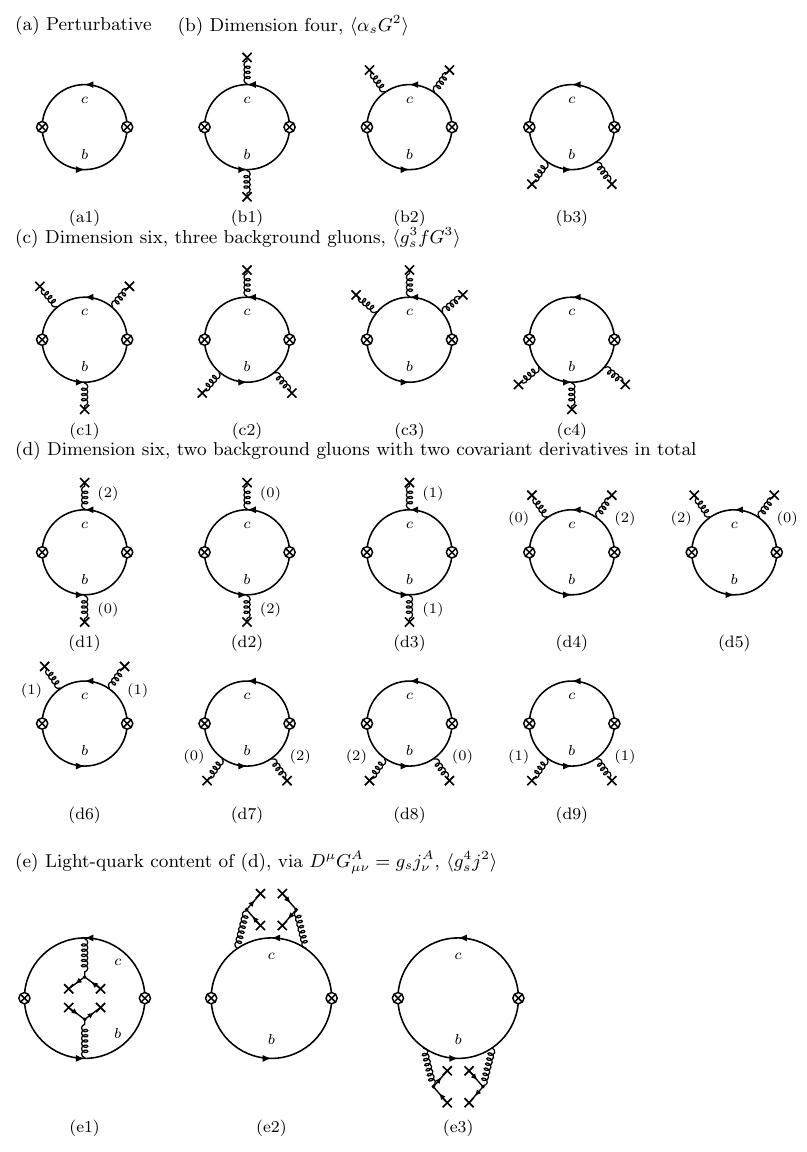}
\caption{Diagrams contributing to the correlation functions up to
dimension six. A label \((r)\) indicates that the gluon field carries \(r\)
covariant derivatives. In diagrams (e), the light-quark lines show the four-quark
content of diagrams (d), which arises through the gluon equation of motion.}
\label{fig:diagrams}
\end{figure}

After vacuum averaging, these propagator terms generate the perturbative,
two-gluon-condensate and three-gluon-condensate parts of the invariant
amplitudes. All diagrams that contribute up to dimension six are shown in
Fig.~\ref{fig:diagrams}. The diagrams (a)--(c) follow from the product of the
two propagators of Eq.~\eqref{eq:heavy_prop_compact}, with the vacuum averages
of the gluon fields given in Ref.~\cite{CostaReinaGubernariWuethrich2026}. The
diagrams (d) require in addition the terms of Eq.~\eqref{eq:fixed_point_gauge}
with covariant derivatives. Through the gluon equation of motion, they
contribute both to \(\langle g_s^3 fG^3\rangle\) and to the four-quark
condensate \(\langle g_s^4 j^2\rangle\), whose light-quark content is shown in
diagrams (e).

After performing Borel transformation, the projected amplitudes can be written as
\begin{equation}
\Pi_1^{ij}(M^2,s_0)
=
\Pi_{1,\mathrm{pert}}^{ij}(M^2,s_0)
+
\Pi_{1,\langle G^2\rangle}^{ij}(M^2,s_0)
+
\Pi_{1,\langle G^3\rangle}^{ij}(M^2,s_0)
{}+
\Pi_{1,\langle g_s^4 j^2\rangle}^{ij}(M^2,s_0),
\qquad i,j=A,B.
\label{eq:full_borel_pi}
\end{equation}
The perturbative contribution is represented by the dispersion integral
\begin{equation}
\Pi_{1,\mathrm{pert}}^{ij}(M^2,s_0)=
\int_{(m_b+m_c)^2}^{s_0} ds\,
\rho_{\mathrm{pert}}^{ij}(s)e^{-s/M^2},
\label{eq:borel_pert}
\end{equation}
where \(\rho_{\mathrm{pert}}^{ij}(s)\) are the spectral densities of the
corresponding correlation functions. They are obtained as
\begin{align}
\rho_{\mathrm{pert}}^{AA}(s)
&=
-\frac{3}{8\pi^2s^2}
\left[(m_b+m_c)^2-s\right]
\left[(m_b-m_c)^2+2s\right]
\sqrt{\lambda(s,m_b,m_c)} ,
\label{eq:rho_AA_pert}
\\
\rho_{\mathrm{pert}}^{AB}(s)
&=
\frac{9}{8\pi^2s(m_b+m_c)}
(m_b-m_c)
\left[(m_b+m_c)^2-s\right]
\sqrt{\lambda(s,m_b,m_c)} ,
\label{eq:rho_AB_pert}
\\
\rho_{\mathrm{pert}}^{BB}(s)
&=
\frac{3}{8\pi^2s(m_b+m_c)^2}
\Big[
-2(m_b^2-m_c^2)^2
+(m_b^2-6m_bm_c+m_c^2)s+s^2
\Big]
\sqrt{\lambda(s,m_b,m_c)} ,
\label{eq:rho_BB_pert}
\end{align}
where
\begin{equation}
\lambda(s,m_b,m_c)=
s^2+m_b^4+m_c^4-2sm_b^2-2sm_c^2-2m_b^2m_c^2 .
\label{eq:kallen}
\end{equation}
Equations~\eqref{eq:rho_AA_pert}--\eqref{eq:rho_BB_pert} are the
leading-order (LO) spectral densities. We also include the NLO
perturbative corrections,
\begin{equation}
\rho^{ij}_{\mathrm{pert}}(s)=\rho^{ij}_{0}(s)+\frac{\alpha_s(\mu)}{\pi}\,
\rho^{ij}_{1}(s,\mu),
\label{eq:rho_nlo}
\end{equation}
where \(\rho^{ij}_0\) are given by
Eqs.~\eqref{eq:rho_AA_pert}--\eqref{eq:rho_BB_pert}. For the diagonal
correlators the \(O(\alpha_s)\) corrections with two massive quarks are known:
\(\rho^{AA}_1\) follows from the vector-current
result~\cite{DjouadiGambino1994,GeneretGubernariLoisa2026} and \(\rho^{BB}_1\)
from the tensor-current result of~\cite{GeneretGubernariLoisa2026}, in
both cases through the replacement \(m_c\to-m_c\), which turns a vector
(tensor) current into an axial-vector (axial-tensor)
one~\cite{DjouadiGambino1994,GeneretGubernariLoisa2026}. The mixed correlator
\(\Pi^{AB}\) has not been calculated before, and we calculate \(\rho^{AB}_1\) in
this work. The one-loop integrals are evaluated with
\textsc{Package-X}~\cite{Patel2015PackageX} through
\textsc{FeynHelpers}~\cite{Shtabovenko2017FeynHelpers}.

The condensate terms are obtained by Feynman parametrizing the corresponding
background-field contributions and applying the Borel transformation directly.
Introducing
\begin{equation}
\bar s(x)=
\frac{m_c^2 x+m_b^2(1-x)}{x(1-x)},
\label{eq:sbar}
\end{equation}
the continuum subtraction restricts the integration region to
\begin{equation}
x_- \leq x \leq x_+,
\qquad
x_\pm =
\frac{s_0+m_b^2-m_c^2 \pm \sqrt{\lambda(s_0,m_b,m_c)}}{2s_0}.
\label{eq:x_region}
\end{equation}

For terms of the form \((Q^2+\bar s)^{-n}\), the Borel transformation gives
\begin{equation}
\mathcal{B}_{M^2}
\left[
\frac{1}{(Q^2+\bar s)^n}
\right]
=
\frac{1}{(n-1)!\,(M^2)^{n-1}}
e^{-\bar s/M^2}.
\label{eq:borel_basic}
\end{equation}
Thus the condensate contributions can be written schematically as
\begin{equation}
\Pi_{1,\langle G^r\rangle}^{ij}(M^2,s_0)
=
\int_{x_-}^{x_+} dx\,
e^{-\bar s(x)/M^2}
\sum_n
\frac{C_{n,r}^{ij}(x)}
{(n-1)!\,(M^2)^{n-1}},
\qquad r=2,3 .
\label{eq:borel_condensates}
\end{equation}
The coefficient functions \(C_{n,r}^{ij}(x)\), including the corresponding
condensate factors for \(r=2,3\), are obtained from the color and Dirac traces
of the background-field propagator contributions after applying the spin-one
projection. Since the explicit expressions are rather lengthy, they are not
displayed here but are available from the authors upon request.
For \(n\geq2\), restricting the \(x\) integration to Eq.~\eqref{eq:x_region}
leaves boundary terms at \(s=s_0\), which we include.

The diagrams (d) lead to the vacuum averages
\(\langle G^A_{\mu\nu}\,D_\rho D_\sigma G^A_{\alpha\beta}\rangle\) and
\(\langle D_\rho G^A_{\mu\nu}\,D_\sigma G^A_{\alpha\beta}\rangle\), which are
expressed through \(\langle g_s^3 fG^3\rangle\) and
\(\langle g_s^4 j^2\rangle\) with the gluon equation of motion,
\(D^\mu G^A_{\mu\nu}=g_s j^A_\nu\), where
\(j^A_\nu=\sum_{q=u,d,s}\bar q\gamma_\nu T^A q\)~\cite{NikolaevRadyushkin1983,CostaReinaGubernariWuethrich2026}.
After the
spin-one projection and Borel transformation, the four-quark contribution takes
the same form as Eq.~\eqref{eq:borel_condensates}. Our four-quark condensate coefficients for
\(\Pi_1^{AA}\) and \(\Pi_1^{BB}\) agree with those
of Ref.~\cite{CostaReinaGubernariWuethrich2026}. In our calculations, we use the vacuum-saturation
approximation
\begin{equation}
\langle g_s^4 j^2\rangle=
-\frac{4}{9}\,r_{\mathrm{vac}}\,g_s^4
\left(\langle\bar uu\rangle^2+\langle\bar dd\rangle^2+\langle\bar ss\rangle^2\right),
\label{eq:jj_vsa}
\end{equation}
where \(r_{\mathrm{vac}}\) parametrizes deviations from
factorization~\cite{Ioffe:2002ee,CostaReinaGubernariWuethrich2026}.

The mixed condensates \(\langle\bar q g_s\sigma Gq\rangle\) and
\(\langle\bar qq\rangle\) do not contribute: the currents contain no light-quark
fields, and since the quark-gluon vertex conserves chirality, such
chirality-odd operators appear only multiplied by a light-quark mass and are
negligible.

Substituting
Eq.~\eqref{eq:full_borel_pi} into
Eq.~\eqref{eq:final_theta_sumrule} gives the final sum rule used in the numerical
analysis.

\section{Numerical analysis}
\label{sec:numerical}

In this section, we present the numerical analysis of the sum rule in
Eq.~\eqref{eq:final_theta_sumrule}. The mixing angle is evaluated as a
function of the Borel parameter \(M^2\) and the continuum threshold \(s_0\).
These auxiliary parameters are varied in a region where the extracted angle
shows a stable behavior, and the uncertainty is estimated by varying both the
QCD input parameters and the sum-rule parameters within the adopted ranges.

The input parameters used in the numerical analysis, together with their
references, are listed in Table~\ref{tab:input_parameters}. The heavy-quark
masses are quoted in the \(\overline{\mathrm{MS}}\) scheme.

\begin{table}[t]
\centering
\caption{Input parameters used in the numerical analysis.}
\label{tab:input_parameters}
\begin{ruledtabular}
\begin{tabular}{lcc}
Parameter & Value & Reference \\
\hline
\(m_b(m_b)\) &
 \((4.186\pm0.006)~\mathrm{GeV}\) &
\cite{ParticleDataGroup:2026aaa} \\

\(m_c(m_c)\) &
 \((1.2729\pm0.0045)~\mathrm{GeV}\) &
\cite{ParticleDataGroup:2026aaa} \\

\(\langle \alpha_s G^2\rangle\) &
\((6.49\pm0.35)\times10^{-2}~\mathrm{GeV}^4\) &
\cite{Narison2021} \\

\(\langle g_s^3 G^3\rangle\) &
\((8.2\pm1.0)~\mathrm{GeV}^2
\langle \alpha_s G^2\rangle\) &
\cite{Narison2021} \\

\(\langle\bar qq\rangle\) (2 GeV), \(q=u,d\) &
\(-(0.278\pm0.022)^3~\mathrm{GeV}^3\) &
\cite{FLAG2021,CostaReinaGubernariWuethrich2026} \\

\(\langle\bar ss\rangle/\langle\bar qq\rangle\) &
\(0.8\pm0.3\) &
\cite{Ioffe:2002ee,CostaReinaGubernariWuethrich2026} \\

\(r_{\mathrm{vac}}\) &
\(1.0\pm0.5\) &
\cite{Ioffe:2002ee,CostaReinaGubernariWuethrich2026} \\

\(\alpha_s(m_Z)\) &
\(0.1180\pm0.0009\) &
\cite{ParticleDataGroup:2026aaa} \\
\end{tabular}
\end{ruledtabular}
\end{table}

For each point in the \((M^2,s_0)\) plane, the invariant amplitudes
\(\Pi_1^{AA}\), \(\Pi_1^{AB}\), and \(\Pi_1^{BB}\) are evaluated from
Eq.~\eqref{eq:full_borel_pi} and substituted into
Eq.~\eqref{eq:final_theta_sumrule}. Since the angle is determined through
\(\tan 2\theta\), the branch of the arctangent is chosen continuously in the
physical region relevant for the \(B_c(1P)\) axial-vector states. The sum rule
contains a ratio of invariant amplitudes; therefore part of the common
normalization uncertainty cancels in the extraction of the angle.

The Borel parameter and continuum threshold must
be chosen in a region where the sum-rule prediction is reliable. We follow the
standard QCD sum-rule criteria: the lower limit of \(M^2\) is fixed by the
convergence of the operator product expansion, while the upper limit is fixed
by requiring that the lowest-state contribution is not dominated by the
continuum. The threshold \(s_0\) is chosen above the squared mass region of
the lowest \(P\)-wave \(B_c\) states and varied around this value to estimate
the residual continuum-threshold dependence. The numerical window is also
consistent with the range used in the QCD sum-rule analysis of vector and
axial-vector \(B_c\) mesons in Ref.~\cite{Wang2013}. In the present analysis,
we use
\begin{equation}
7~\mathrm{GeV}^2 \leq M^2 \leq 9~\mathrm{GeV}^2,
\qquad
53~\mathrm{GeV}^2 \leq s_0 \leq 55~\mathrm{GeV}^2 .
\label{eq:numerical_window}
\end{equation}

The NLO terms are evaluated at the renormalization scale
\(\mu=3~\mathrm{GeV}\), with \(\alpha_s(\mu)\) and the
\(\overline{\mathrm{MS}}\) masses obtained by four-loop running; \(\mu\) is varied
between \(2.5~\mathrm{GeV}\) and \(m_b(m_b)\). The results for three scales are
given in Table~\ref{tab:nlo_scale}.

\begin{table}[t]
\centering
\caption{Running parameters and the mixing angle at LO and NLO for three
values of the renormalization scale, at \(M^2=8~\mathrm{GeV}^2\) and
\(s_0=54~\mathrm{GeV}^2\).}
\label{tab:nlo_scale}
\begin{ruledtabular}
\begin{tabular}{cccccc}
\(\mu\) (GeV) & \(\bar m_b(\mu)\) (GeV) & \(\bar m_c(\mu)\) (GeV) &
\(\alpha_s(\mu)\) & \(\theta^{\mathrm{LO}}\) & \(\theta^{\mathrm{NLO}}\) \\
\hline
2.5 & 4.675 & 1.032 & 0.273 & \(45.8^\circ\) & \(46.8^\circ\) \\
3.0 & 4.481 & 0.989 & 0.253 & \(44.9^\circ\) & \(46.6^\circ\) \\
4.18 & 4.187 & 0.924 & 0.225 & \(43.5^\circ\) & \(47.0^\circ\) \\
\end{tabular}
\end{ruledtabular}
\end{table}

\begin{figure}[t]
\centering
\includegraphics[width=0.82\textwidth]{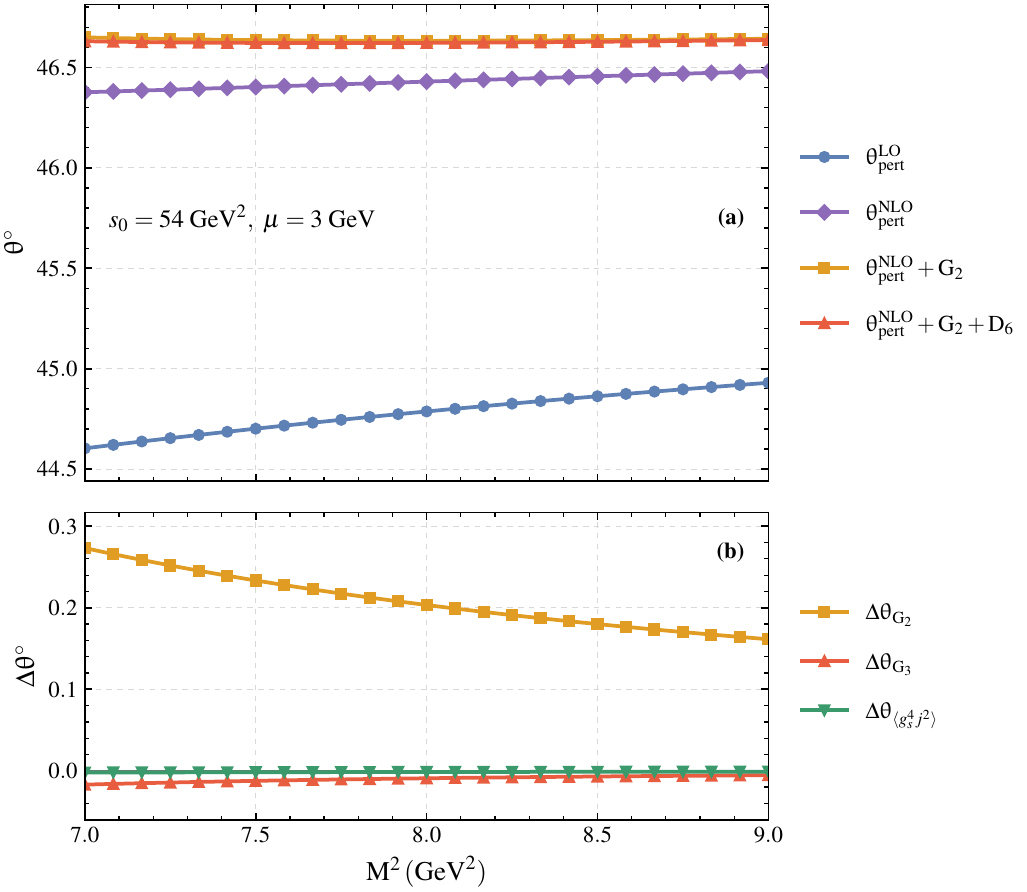}
\caption{Borel-mass dependence of the mixing angle at
\(s_0=54~\mathrm{GeV}^2\) and
\(\mu=3~\mathrm{GeV}\). Panel (a) shows the perturbative result at LO and
at NLO and the successive inclusion of the dimension-four and dimension-six
 contributions; D\(_6\) denotes the three-gluon and
four-quark terms. Panel (b) displays the corresponding incremental shifts,
taken relative to the NLO perturbative result,
\(\Delta\theta_{G_2}=\theta_{\mathrm{pert}+G_2}-\theta_{\mathrm{pert}}\),
\(\Delta\theta_{G_3}=\theta_{\mathrm{pert}+G_2+G_3}
-\theta_{\mathrm{pert}+G_2}\), and the four-quark shift
\(\Delta\theta_{\langle g_s^4j^2\rangle}\).}
\label{fig:ope_diagnostics}
\end{figure}

In Fig.~\ref{fig:ope_diagnostics}, we present the dependence of the mixing angle on \(M^2\) at \(s_0=54~\mathrm{GeV}^2\) and \(\mu=3~\mathrm{GeV}\). From this figure, we deduce that the perturbative contribution gives
the main part of the result, while adding the two-gluon and three-gluon
condensate terms produces only a tiny shift in the extracted angle.
At \(M^2=8~\mathrm{GeV}^2\), the NLO corrections increase the angle by
\(1.6^\circ\), the two-gluon condensate by \(0.2^\circ\), and the
dimension-six terms change it by \(-0.01^\circ\).

In the three-gluon term, the contributions of the diagrams (c3), (c4), and
(d) largely cancel; our three-gluon coefficients for \(\Pi_1^{AA}\) and
\(\Pi_1^{BB}\) agree with those of Ref.~\cite{CostaReinaGubernariWuethrich2026} up to
the higher-order terms of the \(m_c\) expansion used there. For the four-quark
condensate, we use \(g_s^4\langle\bar qq\rangle^2(\mu)=16\pi^2\alpha_s(\mu)\,
[\alpha_s\langle\bar qq\rangle^2]_{2~\mathrm{GeV}}\), which with the inputs of
Table~\ref{tab:input_parameters} and \(r_{\mathrm{vac}}=1\) gives
\(\langle g_s^4 j^2\rangle=-6.5\times10^{-3}~\mathrm{GeV}^6\) at
\(\mu=3~\mathrm{GeV}\); this term changes the mixing angle by only
\(-0.001^\circ\).

\begin{figure}[t]
\centering
\includegraphics[width=0.82\textwidth]{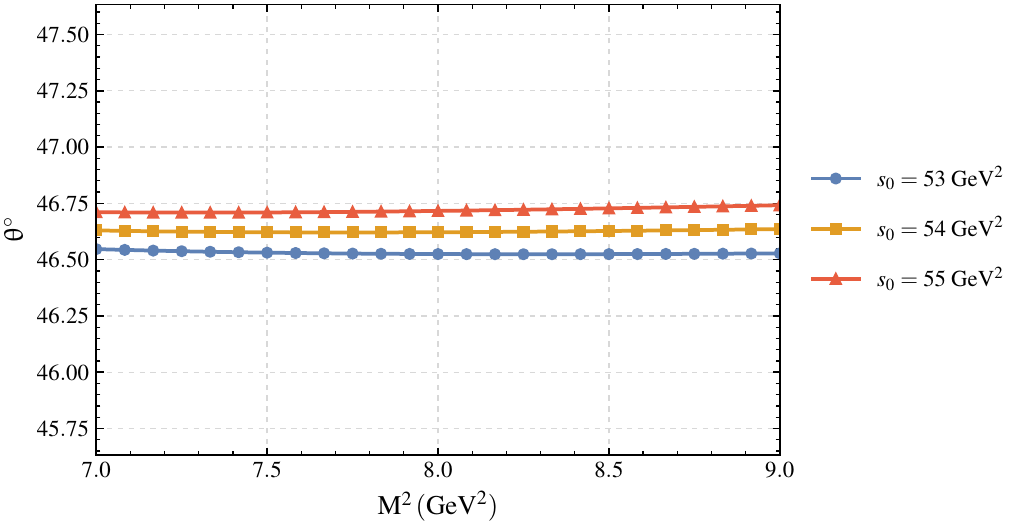}

\vspace{0.4cm}

\includegraphics[width=0.82\textwidth]{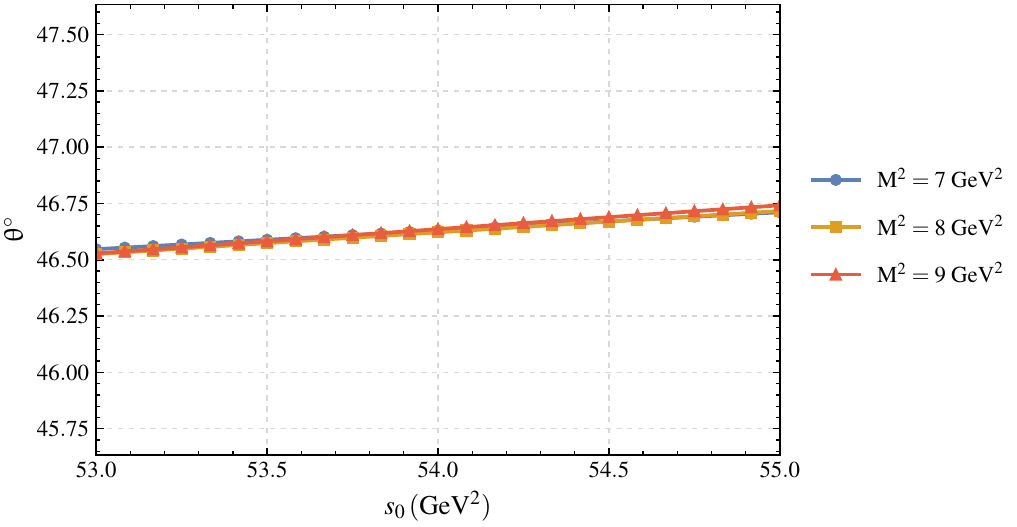}
\caption{Dependence of the extracted \(B_c(1P)\) axial-vector mixing angle on
the auxiliary parameters in the selected sum-rule window, at NLO and
\(\mu=3~\mathrm{GeV}\). Upper panel:
Dependence on the Borel parameter \(M^2\) for different fixed values of the
continuum threshold \(s_0\). Lower panel: Dependence on \(s_0\) for different
fixed values of \(M^2\).}
\label{fig:theta_stability}
\end{figure}

\begin{figure}[t]
\centering
\includegraphics[width=0.82\textwidth]{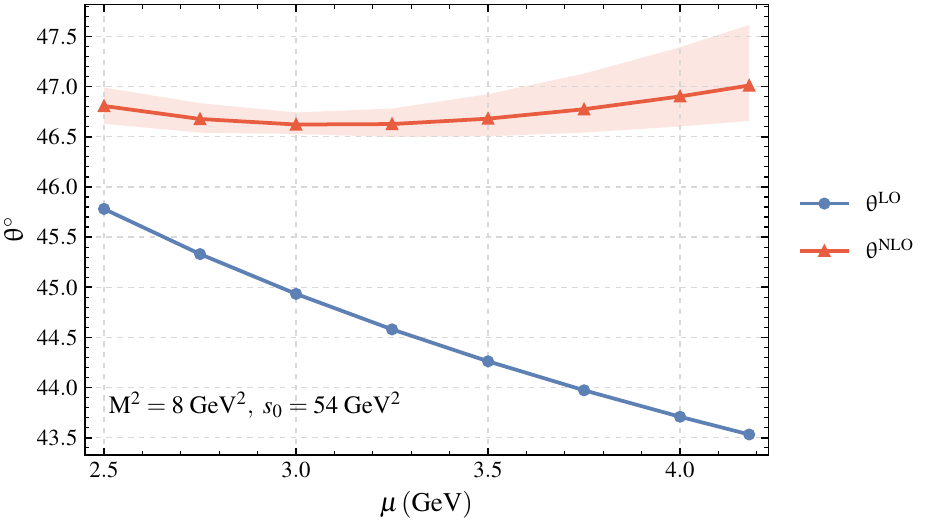}
\caption{Dependence of the mixing angle on the renormalization scale
\(\mu\) at LO and NLO, at \(M^2=8~\mathrm{GeV}^2\) and \(s_0=54~\mathrm{GeV}^2\).
The band shows the variation of the NLO result within the window of
Eq.~\eqref{eq:numerical_window}.}
\label{fig:theta_mu}
\end{figure}

In the chosen window, the physical result should be almost independent of the
auxiliary parameters. This stability is checked explicitly in
Fig.~\ref{fig:theta_stability}, with the NLO corrections included and
\(\mu=3~\mathrm{GeV}\). The upper panel shows the dependence of the
mixing angle on \(M^2\) for fixed values of \(s_0\), while the lower panel
shows the dependence on \(s_0\) for fixed values of \(M^2\). In both cases,
the extracted angle changes only weakly in the selected working region.
The scale dependence is shown in Fig.~\ref{fig:theta_mu}.

To estimate the uncertainty, we vary \(\mu\), \(M^2\), \(s_0\), and the inputs
of Table~\ref{tab:input_parameters} within their ranges, one at a time, around
\(\mu=3~\mathrm{GeV}\), \(M^2=8~\mathrm{GeV}^2\), and \(s_0=54~\mathrm{GeV}^2\).
The resulting shifts are listed in Table~\ref{tab:budget}; adding them in
quadrature,
\begin{table}[t]
\centering
\caption{Uncertainties of the mixing angle from the individual sources.}
\label{tab:budget}
\begin{ruledtabular}
\begin{tabular}{lc}
Source & \(\Delta\theta\) \\
\hline
Scale, \(2.5~\mathrm{GeV}\leq\mu\leq m_b(m_b)\) & \(0.39^\circ\) \\
Continuum threshold, \(53\)--\(55~\mathrm{GeV}^2\) & \(0.10^\circ\) \\
\(\alpha_s(m_Z)\) & \(0.05^\circ\) \\
\(m_c(m_c)\) & \(0.025^\circ\) \\
Borel parameter, \(7\)--\(9~\mathrm{GeV}^2\) & \(0.014^\circ\) \\
\(\langle\alpha_sG^2\rangle\) & \(0.011^\circ\) \\
\(m_b(m_b)\) & \(0.007^\circ\) \\
Four-quark condensate & \(0.003^\circ\) \\
\(\langle g_s^3G^3\rangle\) & \(0.001^\circ\) \\
\hline
Total (in quadrature) & \(0.41^\circ\) \\
\end{tabular}
\end{ruledtabular}
\end{table}
this procedure gives
\begin{equation}
\theta_{B_c(1P)}=(46.6\pm0.4)^\circ .
\label{eq:theta_final}
\end{equation}
The quoted uncertainty includes the variation of the renormalization
scale, Borel parameter,
continuum threshold, heavy-quark masses, \(\alpha_s\), and condensate inputs. This value indicates a nearly maximal mixing between the \(1^1P_1\) and \(1^3P_1\)
components of the physical \(B_c(1P)\) axial-vector states. In other words,
the physical \(1^+\) states are expected to be strong mixtures of the two
axial-vector configurations rather than nearly pure \(1^1P_1\) or
\(1^3P_1\) states.

Our numerical analysis shows that the NLO corrections raise the angle by about \(1.6^\circ\). The resulting angle is close to \(45^\circ\), so the
two physical axial-vector states are nearly equal mixtures of the \(1^1P_1\) and
\(1^3P_1\) configurations.

The \(B_c(1P)\) axial-vector mixing angle has been studied in several
theoretical approaches. A comparison of representative results is given in
Table~\ref{tab:theta_comparison}. The predictions show a sizable model
dependence. Most positive values lie in the range
\(17^\circ\!\!-\!36^\circ\), while some analyses give larger or
convention-dependent results, extending up to about \(55^\circ\). Our QCD
sum-rule result lies within the broad spectrum of existing theoretical
predictions.

\begin{table}[t]
\centering
\caption{Comparison of the \(1^1P_1-1^3P_1\) mixing angle of the
\(B_c(1P)\) axial-vector states obtained in different theoretical approaches.
The values quoted from the literature follow the convention used in the
corresponding references.}
\label{tab:theta_comparison}
\begin{ruledtabular}
\begin{tabular}{lc}
Method / Model & \(\theta\) \\
\hline
Lattice QCD \cite{Davies1996} & \(33.4^\circ\) \\
Potential model \cite{Gershtein1995} & \(17.1^\circ\) \\
QCD potential model \cite{GuptaJohnson1996} & \(25.6^\circ\) \\
Phenomenological model \cite{Fulcher1999} & \(28.5^\circ\) \\
Relativistic quark model \cite{Ebert2003} & \(20.4^\circ\) \\
Relativized quark model \cite{Godfrey2004} & \(22.4^\circ\) \\
Quark model \cite{EichtenQuigg2019} & \(18.7^\circ\) \\
Nonrelativistic quark model \cite{Li2019} & \(35.5^\circ\) \\
Bethe-Salpeter approach \cite{Wang2022} & \(32.2^\circ\) \\
Modified Godfrey-Isgur model \cite{Li2023} & \(-24.3^\circ\) \\
Spectroscopic survey \cite{LiLiu2023} & \(35.2^\circ\) \\
Coupled-channel model \cite{HaoZhu2024} & \(55.0^\circ\) \\
QCD sum rules (This work) &  \((46.6\pm0.4)^\circ\) \\
\end{tabular}
\end{ruledtabular}
\end{table}

The phenomenological importance of this angle is enhanced by the recent LHCb
observation of a broad peaking structure compatible with the lowest
\(B_c(1P)^+\) multiplet in the \(B_c^+\gamma\) spectrum
\cite{LHCb2025PRL,LHCb2025PRD}. Since the individual contributions from the
members of the \(1P\) multiplet have not yet been resolved, theoretical
information on the masses, radiative transition patterns, and axial-vector
mixing angle is important for assigning the observed structures and for
constructing theory-constrained fit models. The mixing angle enters the
radiative transition amplitudes of the axial-vector states and can affect the
relative contributions of the different \(P\)-wave states to the observed
\(B_c^+\gamma\) signal. As larger data samples and more detailed studies of the
excited beauty-charm spectrum become available, independent nonperturbative
inputs such as the present QCD sum-rule determination can help constrain
phenomenological models and guide future interpretations of the \(B_c(1P)\)
sector.

\section{Conclusion}
\label{sec:conclusion}

We have determined the \(1^1P_1-1^3P_1\) mixing angle in the \(B_c(1P)\)
axial-vector sector using QCD sum rules. The analysis was based on two-point
correlation functions constructed from axial-vector and tensor interpolating
currents and evaluated using the operator product expansion. Perturbative,
two-gluon-condensate, and three-gluon-condensate contributions were included,
together with the four-quark condensate and the next-to-leading-order
perturbative corrections, which we calculated for the mixed correlator of the
axial-vector and tensor currents. The NLO corrections raise the angle by about
\(1.6^\circ\) and reduce its scale dependence, while the dimension-six condensate
contributions are negligible.

After Borel transformation and continuum subtraction, the mixing angle was
extracted from the diagonalization condition of the projected spin-one
correlation matrix. The final result is
\begin{equation}
\theta_{B_c(1P)} = (46.6\pm0.4)^\circ .
\end{equation}
This value indicates a nearly maximal mixing of the two axial-vector
configurations \(1^1P_1\) and \(1^3P_1\). We also perform comparison of our results with existing theoretical predictions on the mixing angle.

The result provides complementary theoretical input for studies of the
recently observed orbitally excited \(B_c^+\) structures in the
\(B_c^+\gamma\) spectrum. As future experimental analyses further resolve the
\(B_c(1P)\) multiplet, the mixing angle can be used in phenomenological studies
of radiative transitions and in theory-constrained descriptions of the excited
beauty-charm spectrum.

\begin{acknowledgments}
We thank N.~Gubernari for useful correspondence regarding
NLO corrections for \(B_c\) systems.
\end{acknowledgments}

\section*{Data availability}
There are no publicly available research data or software supporting this
manuscript. Requests for further information or data should be sent to the
authors.

\bibliographystyle{utcaps_mod}
\bibliography{Bc_mixing}


\end{document}